\documentclass[12pt,amsmath,amssymb,prd,superscriptaddress]{revtex4}

\usepackage{color}
\usepackage{graphicx}
\usepackage{isotope}
\usepackage{siunitx}
\newcommand{\beqn}{\begin{equation}}
\newcommand{\eeqn}{\end{equation}}

\newcommand{\fig}[1]{Fig.\,\ref{fig:#1}}

\begin{document}

\title{Laser-driven neutron source from high temperature D-D fusion reactions}
\author{X. Jiao}
\affiliation{Center for High Energy Density Science, University of Texas at Austin, Austin, Texas 78712, USA}
\author{C. Curry}
\affiliation{SLAC National Accelerator Laboratory, Menlo Park, California 94025, USA}
\author{M. Gauthier}
\affiliation{SLAC National Accelerator Laboratory, Menlo Park, California 94025, USA}
\author{F. Fiuza}
\affiliation{SLAC National Accelerator Laboratory, Menlo Park, California 94025, USA}
\author{J. Kim}
\affiliation{SLAC National Accelerator Laboratory, Menlo Park, California 94025, USA}
\author{E. McCary}
\affiliation{Center for High Energy Density Science, University of Texas at Austin, Austin, Texas 78712, USA}
\author{L. Labun}
\affiliation{Center for High Energy Density Science, University of Texas at Austin, Austin, Texas 78712, USA}
\author{O. Z. Labun}
\affiliation{Center for High Energy Density Science, University of Texas at Austin, Austin, Texas 78712, USA}
\author{C. Schoenwaelder}
\affiliation{SLAC National Accelerator Laboratory, Menlo Park, California 94025, USA}
\author{R. Roycroft}
\affiliation{Center for High Energy Density Science, University of Texas at Austin, Austin, Texas 78712, USA}
\author{G. Tiwari}
\affiliation{Center for High Energy Density Science, University of Texas at Austin, Austin, Texas 78712, USA}
\author{G. Glenn}
\affiliation{Center for High Energy Density Science, University of Texas at Austin, Austin, Texas 78712, USA}
\author{F. Treffert}
\affiliation{SLAC National Accelerator Laboratory, Menlo Park, California 94025, USA}
\author{T. Ditmire}
\affiliation{Center for High Energy Density Science, University of Texas at Austin, Austin, Texas 78712, USA}
\author{S. Glenzer}
\affiliation{SLAC National Accelerator Laboratory, Menlo Park, California 94025, USA}
\author{B. M. Hegelich}
\affiliation{Center for High Energy Density Science, University of Texas at Austin, Austin, Texas 78712, USA}

\date{16 October 2020}

\begin{abstract}
We report a laser-driven neutron source with high yield ($>10^8$/J) and high peak flux ($>10^{25}$\si{\per\square\cm\per\second}) derived from high-temperature deuteron-deuteron fusion reactions. The neutron yield and the fusion temperature ($\sim 200$ keV) in our experiment are respectively two orders of magnitude and one order of magnitude higher than any previous laser-induced D-D fusion reaction. The high-temperature plasma is generated from thin ($\sim 2$\si{\micro\meter}), solid-density deuterium targets, produced by a cryogenic jet, irradiated by a 140 fs, 130 J petawatt laser with an F/3 off-axis parabola and a plasma mirror achieving fast volumetric heating of the target.  The fusion temperature and neutron fluxes achieved here suggest future laser experiments can take advantage of neutrons to diagnose the plasma conditions and come closer to laboratory study of astrophysically-relevant nuclear physics.
\end{abstract}

\maketitle

\section{Introduction}
Laser-driven nuclear fusion remains an active research topic, aimed at generating either alternative energy \cite{hurricane2014fuel} or an intense neutron flux from a compact, controllable source \cite{taylor2007route}.  Much work so far as focused on inertial confinement fusion (ICF) \cite{glenzer2010symmetric}, in which lasers compress fusion fuels, directly \cite{soures1996direct} or indirectly \cite{betti2016inertial}, heating it to several keV \cite{edwards2013progress} through plasma shockwaves, with the goal of initiating self-sustaining fusion reaction and thus generating energy. While these experiments can provide high neutron number per shot that can be used for specific fundamental physics studies, ICF user facilities, especially Omega \cite{boehly1997initial} or NIF \cite{spaeth2016description}, are too large to be used as a more general flexible neutron source. 

The advent of ultrahigh intensity lasers using Chirped Pulse Amplification \cite{strickland1985compression} has changed this paradigm and enables high-flux neutron sources at a practical neutron-per-Joule efficiency in a much more compact setup.  High-intensity lasers have the potential to satisfy the need for compact and intense neutron sources in research fields and industry, including bulk temperature measurements in dynamic materials and high energy density physics \cite{yuan2005shock}, material science \cite{higginson2010laser}, medical research \cite{barth2005boron}, and astrophysics \cite{wallerstein1997synthesis}. 

Several schemes have been tested on laser systems  ranging in scale from millijoule \cite{hah2016high} to kilojoule \cite{higginson2010laser}. Among them, the most promising method is the ion-driven neutron source, because of its comparatively high efficiency and yield. Experiments have demonstrated a neutron yield of $>10^{10}$/sr/shot \cite{roth2013bright}. The scheme is usually employed in a ``pitcher-catcher'' configuration, where the laser interacts with a thin (few $\mu$m-thick) target, or ``pitcher'', to accelerate ions, typically protons or better yet deuterons. The accelerated ions then interact with a cm-scale, low-Z metal ``catcher'', often beryllium or lithium \cite{higginson2011production}, undergoing nuclear reactions and producing neutrons in the process.  In this scheme, the overall neutron-to-laser-energy efficiency depends on the laser-ion acceleration efficiency.  Break-out afterburner acceleration in the relativistic transparency regime, the most efficient demonstrated laser-ion acceleration mechanism \cite{yin20113dboasims,hegelich2013laser}, achieves a neutron number-to-laser energy conversion efficiency $>10^8$/J \cite{roth2013bright}.  In contrast, TNSA acceleration of ions has achieved $\sim 10^7$/J, a laser-electron driven neutron source $\sim 10^6$/J \cite{jiao2017tabletop}, and laser-irradiated clusters only $\sim 10^5$/J \cite{bang2013experimental}.
 
One of the drawbacks of the pitcher-catcher approach is the much greater size of the neutron source ($\sim$ 10\si{\centi\meter\cubed}) compared to the ion source ($\sim$100-1000\si{\micro\meter\cubed}) due to the catcher size and ion stopping ranges in the converter. The greater source size reduces the neutron surface flux by $\sim (10^2)^2$ compared to the maximum achievable by smaller ion sources.

In this letter, we report a novel approach to laser-neutron generation aimed at addressing this shortcoming. For the first time in a petawatt laser facility, pure deuterium ice targets were tested for their potential in neutron generation. These targets offer several advantages over the deuterated plastic foils used in previous experiments: electron density tuned to the relativistic transparency  regime \cite{palaniyappan2012dynamics} with PW laser parameters; high purity and therefore higher efficiency and fewer loss channels; and compability with high laser repetition rate \cite{gauthier2017high}. The neutron production scheme is based on volumetrically heating deuterons inside a relativistically-transparent target at solid density. The deuterons were heated to $\gtrsim 100$ keV, triggering high D-D fusion reaction rates in a volume much smaller than the typical converter size, thus leading to a very high neutron flux $> 10^{19}$\si{\per\centi\meter\squared\per\second} at 1\,cm from the reaction region.  The neutrons are estimated to be delivered over an approximately 10 ps plasma confinement time, corresponding to a neutron density $\gtrsim 10^{9}$\si{\per\centi\meter\squared}.

\section{Experimental setup}
\begin{figure}
\includegraphics[width=\textwidth]{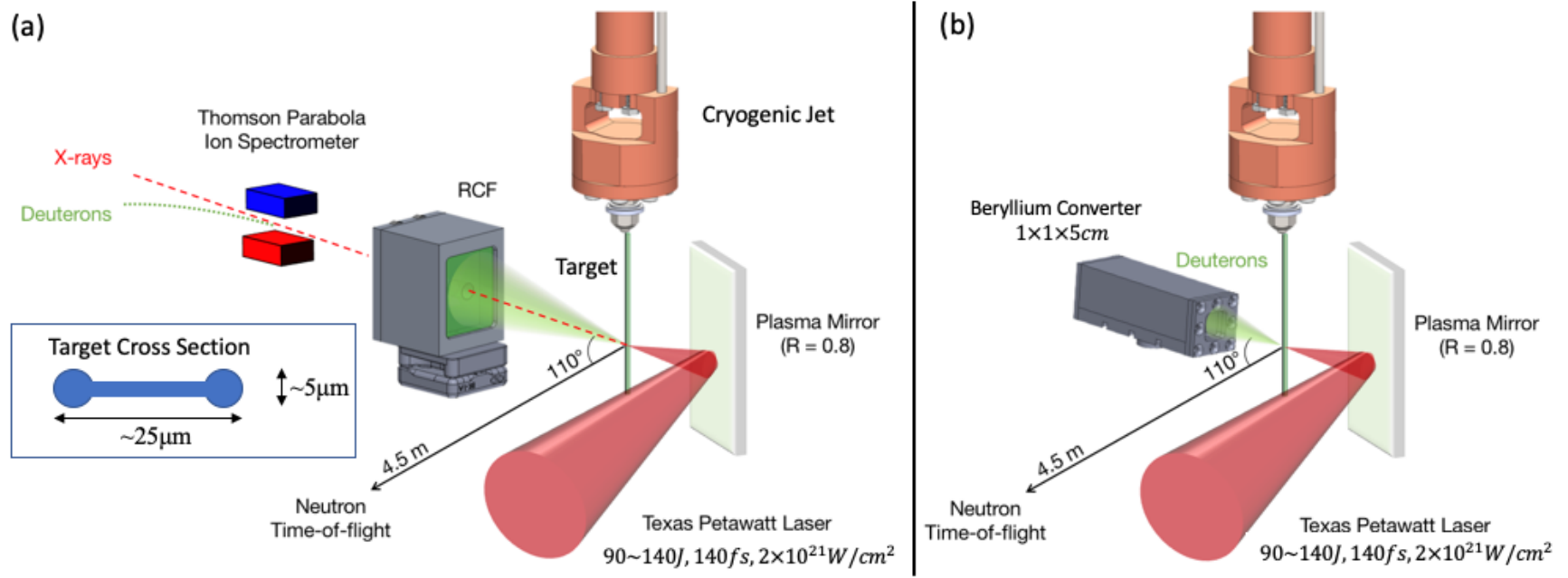}
\caption{ a) Experiment setup with RCF stack. Inset: cross section of the deuterium ice targets. b) Experiment setup with beryllium converter. 
\label{fig:experimentalsetup}}
\end{figure}

The experiment was carried out at the Texas Petawatt laser \cite{gaul2010demonstration}  facility at the University of Texas at Austin. The experimental setup is depicted in \fig{experimentalsetup}. The 1057 nm Nd:Glass laser delivered 90-140 J, 140-fs laser pulses to irradiate the deuterium ice targets. The laser pulse was focused with an F/3 final optic to a spot size of 6\si{\micro\meter} (FWHM) providing an average encircled laser intensity $2\times 10^{21}$ W/cm$^2$. A plasma mirror was installed 5 cm before the target to remove pre-pulses and steepen the rise of the main pulse, achieving a contrast ratio $>10^5$ at 10 ps before the arrival of the peak (see \fig{TPWprofile} in the appendix).  Measurements indicate the plasma mirror reflects approximately 80\% of the laser energy.

The deuterium ice target was generated by a cryogenic microjet system \cite{gauthier2016high}. Deuterium was cooled to 17 K in liquid form by a continuous-flow helium cryostat. The liquid deuterium solidified with little mass loss after evaporating through a nozzle with a $2\times 40\,\si{\micro\meter}$ rectangular opening. The resulting deuterium ice target is a relatively flat sheet $\sim 15\,\si{\micro\meter}$ wide, 2\,\si{\micro\meter} thick with $\simeq 5$\si{\micro\meter} diameter cylindrical columns at each end, as illustrated in the inset of \fig{experimentalsetup}a, running at $50-100$ m/s into vacuum. The laser irradiated the flat part of the target at an angle of 30 degrees from the sheet normal. 

Either a radiochromic film (RCF) stack or a beryllium (Be) converter was placed on the laser propagation axis through the target ($30^\circ$ from target normal). The RCF stack consisted of alternating square-shape layers of aluminum foil, copper sheets and calibrated RCF films of different thickness to measure the deuteron beam divergence and to estimate the spectrum. The precise composition of the stack is shown in the supplemental material. The RCF stack was placed 4.5 cm behind the target and contains a pinhole at the center point to pass a narrow central section of the beam uninhibited to the Thomson parabola spectrometer (TPS), which provided a higher resolution deuteron spectrum measurement at the laser direction compared to the RCF stack. The RCF stack also served as a neutron converter where deuterons produced neutrons via breakup reactions. Alternatively to the RCF stack, we placed a berrylium block 2.7 cm behind the target, which yielded a higher neutron conversion efficiency due to its low atomic weight. The beryllium converter had dimensions of $1\times 1\times 5$ cm and was placed inside an aluminum can with the frontside covered by $10\mu$m-thick aluminum foil to mitigate the production of potentially harmful beryllium dust by the laser pulse. 

A neutron time of flight (n-TOF) detector measured the neutron energy spectrum 4.5m away from the target chamber center (TCC) and 110 degrees away from the laser propagation direction in the laser polarization plane, which was the only position available due to the radiation shielding. The n-TOF consisted of a fast plastic scintillator (EJ-200), a photomultiplier (XP2020) and a fast-digital oscilloscope (TDS5014). The strong x-ray signal from the laser-target interaction was suppressed by a 32.5mm-thick Cu plate to limit the decay signal width (FWHM) to $\leq 25$ ns, so that it would not overwhelm the neutron signal but remain strong enough to trigger the n-TOF system and serve as a time reference for the neutron energy analysis. The scintillator response function was measured beforehand, and the width (FWHM) was found to be around 10ns per volt.  This width introduces $\sim 10\%$ uncertainty in the energy measurement, affecting higher energies more than lower energies. Ten bubble detectors \cite{ing1997bubble} were positioned around and on top of the target chamber to capture neutrons escaping from the target and the converter. The bubble detectors have sensitivities of $\sim 2$ bubbles/mrem and were individually calibrated against a radioactive neutron source before the campaign. 

\section{Results and analysis}
\begin{figure}[b]
\includegraphics[width=0.5\textwidth]{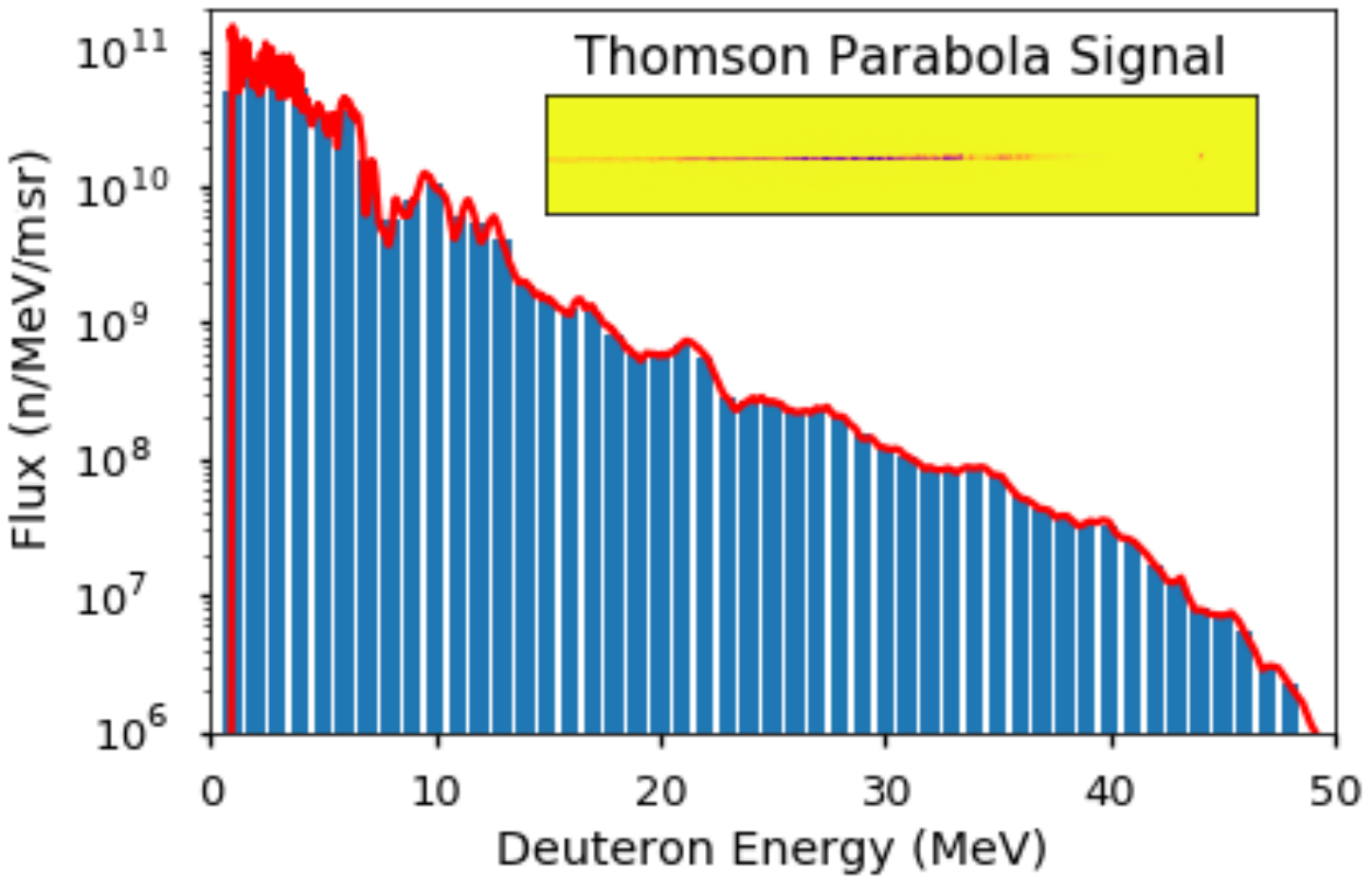}
\caption{ Deuteron spectrum from the laser plasma acceleration. Inset: TPS image of the deuteron trace. 
\label{fig:dspectrum}}
\end{figure}

The deuteron energy spectrum in laser direction was measured by the TPS. The absolute flux was calculated using the image plate calibration from Ref. \cite{alejo2014characterisation}. The deuteron spectrum was found to follow an exponential distribution with energy up to 50 MeV, as shown in \fig{dspectrum}. The high voltage setting on the TPS was disabled to avoid electrical arcing due to the continuous presence of evaporating gas inside the chamber.  A previous experiment showed that all other ion species are negligible compared to deuterons from the jet because of the high purity of the source \cite{gauthier2016high}. 

\begin{figure}[t]
\includegraphics[width=0.5\textwidth]{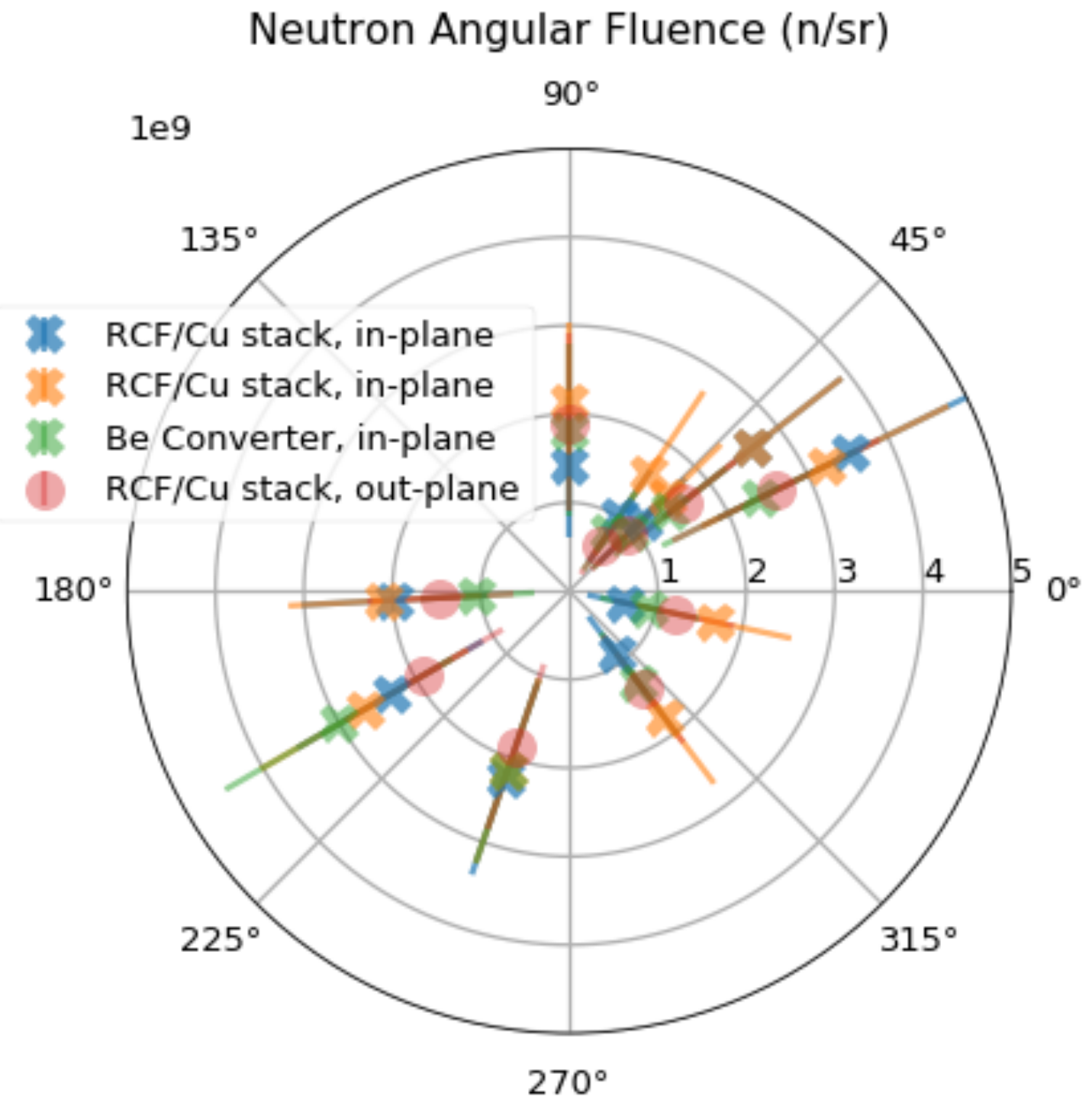}
\caption{ Neutron angular fluence measured in-plane (horizontal) or out-plane at different directions with either RCF stack or beryllium converter.  $0^\circ$ is the laser direction.
\label{fig:neutronbubble}}
\end{figure}

The total neutron yield was found to be roughly $2\times 10^{10}$/shot consistently across all successful shots.  We estimated the total yield from the fluxes measured by bubble detectors surrounding the target chamber.  Absolute neutron number was determined by converting bubble counts to dosage (measured in mrem) immediately after the detectors' irradiation, and the dosages were converted to absolute neutron numbers using the relation provided in Ref. \cite{jung2013characterization}. Neutron fluences are not isotropic but every shot shows a consistent pattern with the highest flux observed near target normal (30$^\circ$ from laser direction), as shown in \fig{neutronbubble}.

\begin{figure}
\includegraphics[width=\textwidth]{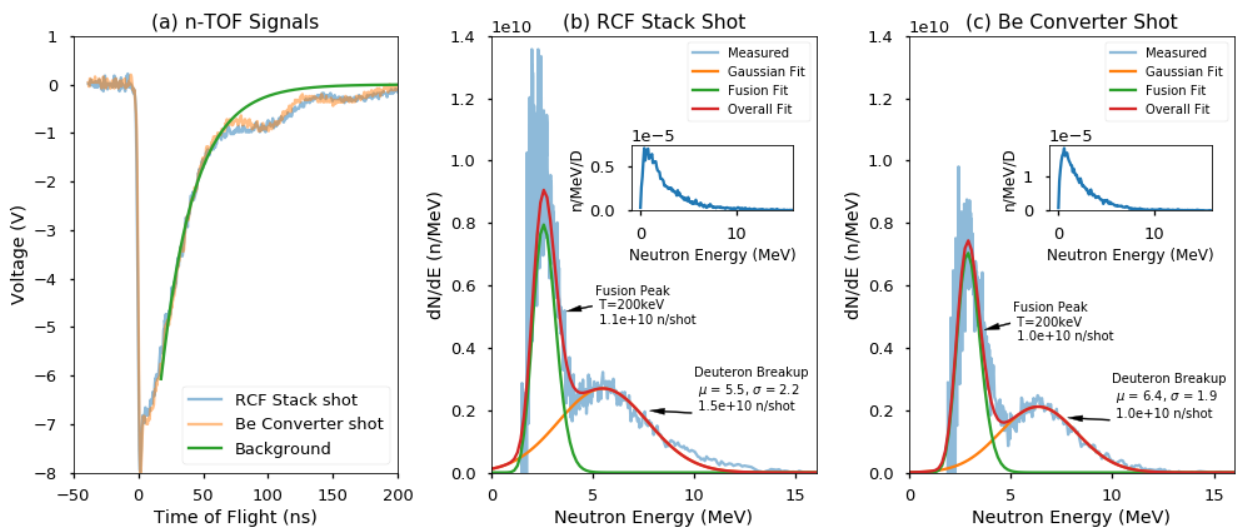}
\caption{a) n-TOF signal and fit to the x-ray fluoresence background. b) Neutron spectrum for shot with the RCF stack. Laser energy 120J ($\simeq 96$J on target after plasma mirror) and total neutron yield $2.6\times 10^{10}$. c) Neutron spectrum for shot with Be converter equipped. Laser energy 93J ($\simeq 75$J on target), total neutron yield is $2.0\times 10^{10}$. Insets: neutron spectrum from deuteron breakup in the RCF stack and Be converter respectively as predicted by GEANT4 simulation. The spectrum is normalized to the incident deuteron number, hence the units MeV$^{-1}$D$^{-1}$.
\label{fig:neutronspectra}}
\end{figure}

Figure \ref{fig:neutronspectra} shows the n-TOF scintillator signal and derived neutron spectra. The spectra were extracted by first subtracting the X-ray signal from the total scintillator signal, which is accomplished by fitting the early-time X-ray-triggered fluorescence with an exponential decay model \cite{jung2013characterization}.  In fitting the X-ray-signal, we exclude the first 15 ns to avoid the interference of the fast decay (non-fluorescent) process, visible as the sharp peaks at time zero in \fig{neutronspectra}(a).  The neutron component of the signal was then converted from time domain to energy domain $dS/dt \mapsto dS/dE$. Finally, signal amplitude $dS$ was converted to neutron number $dN$ by dividing by a theoretically-calculated calibration curve ($dS/dN$) for each energy bin as described by Ref. \cite{o1996response}.

The resulting neutron spectrum from shots with RCF stack and Be converter are shown in \fig{neutronspectra} b and c. We identified two distinct peaks: one narrower and peaked around 2.8 MeV, the other broader and peaked around 6 MeV, indicating two different sources of neutrons. The peak at 2.8 MeV was fit well by the distribution expected from the $d+d\to n+\isotope[3]{He}$ reaction in a thermal plasma \cite{appelbe2011production} with the deuteron distribution described by a temperature $T_d=200$ keV and no bulk flow velocity, $\vec v_b=0$.  Due to noise in the neutron spectrum at low energy (late times), the fit has a relatively large uncertainty of $50$\,\si{\kilo\electronvolt}.

With the $2.8$\,MeV peak identified as sourced from fusion reactions, the broader distribution peaked at higher energy stems from deuteron breakup reactions and (d,n) reactions in either the RCF/Al/Cu stack or the beryllium converter.  The deuteron break-up cross section is not well measured for most materials.  However, the break-up reaction is kinematically suppressed for center of mass energy below the 2.2 MeV binding energy of the deuteron.  Therefore, the breakup cross-section must decrease with deuteron energy, and the break-up reaction's contribution to the neutron spectrum should also decrease as the neutron energy approaches zero.  We find that the second peak is well-described by a gaussian function.  
 
To predict the break-up contribution to the neutron spectrum, we simulated the deuteron beam interaction with the converter using theoretical cross sections for deuteron breakup built into GEANT4 \cite{allison2016recent}. We modeled the deuteron beam with the measured energy spectrum (\fig{dspectrum}) interacting with a beryllium converter and an RCF stack in the experimental configuration (\fig{experimentalsetup}).  Since the TPS was on the same axis as the converter, we expect the TPS-derived spectrum to be a reasonable sample of the deuteron distribution entering the converters. The resulting spectra are shown in the insets of \fig{neutronspectra}. GEANT4 predicts that neutrons emitted from the RCF stack and the Be converter exhibit almost the same spectral profile with energies up to 16 MeV and a single peak structure. The neutron conversion efficiency is almost two times higher in the Be converter than in the RCF stack.  The lower neutron yield derived from break-up reactions in the Be shot must be attributed to lower deuteron flux due in turn to $\sim 20\%$ lower on-target laser energy. For the RCF stack, for which an estimate of total deuteron number is available from the spectrum, the number of break-up neutrons predicted to reach the TOF detector is consistent with the measured value of $1.5\times 10^{10}$ at the order of magnitude level.  However, the predicted spectrum does not match the observed spectrum, likely due to a combination of limited knowledge (uncertain cross sections and off-axis deuteron distribution) and simplifications in modeling the environment (e.g. additional structures between the converter and the detector).  As the spectral shape depends on the deuteron spectrum, beam divergence, incidence angle, deuteron break-up model and environmental scattering, exact replication is difficult to achieve with the limited amount of data available.

Combining the spectral data from n-TOF detector and the neutron dose from the bubble detectors, we found that the fusion neutron yield is close to $1\times 10^{10}$/shot for all successful shots, corresponding to $\simeq 10^8$/J neutron-per-laser-energy conversion efficiency. 

From the total yield and theoretically calculated neutron production rate of $4\times 10^{35}$/m$^3$/s in a $T_d\simeq 200$\,keV, $n_d\simeq 6\times 10^{28}$/m$^3$ plasma, we can estimate the reaction volume \cite{pfalzner2006introduction}.  Assuming a plasma confinement time of 10 ps, the reaction volume is approximately 3000\si{\micro\meter\cubed}, corresponding roughly to a plasma sphere of 9\,\si{\micro\meter} radius, similar in scale to the target and laser focus dimensions.  Because the d-d reaction rate decreases as $1/R^6$ where $R$ is the size scale of the plasma, the observed high yield and zero-bulk velocity fit to the neutron spectrum suggest that most fusion neutrons originated at early times in this reaction volume before significant expansion and temperature change.  Zero bulk velocity also shows that Coulomb repulsion from charge imbalance in the plasma had no significant impact on the plasma dynamics, which is consistent with the expectation that the dimensions of the reaction volume are much larger than the skin depth so that only a very small portion of the electrons escape from the surface of the plasma.

Considering the Texas Petawatt laser with focal spot size of 6\si{\micro\meter} (FWHM) and incidence angle of 30$^\circ$, we independently estimate the dimensions reaction volume as being 10s of \si{\micro\meter} transversely.  Taking the total neutron number in the fusion peak $N_{n,{\rm fusion}}\simeq 1\times 10^{10}$ and dividing by a conservative confinement time $\tau=10$\,\si{\pico\second}, we obtain neutron flux values of $8\times 10^{25}$\si{\per\second\per\centi\meter\squared} on the surface of the reaction volume represented by a 10\,\si{\micro\meter}-radius sphere.  The fusion neutron flux drops to $8\times 10^{23}$\,\si{\per\second\per\centi\meter\squared} at 100\,\si{\micro\meter} from the reaction volume and $8\times 10^{19}$\,\si{\per\second\per\centi\meter\squared} at 1\,\si{\centi\meter}, distances at which secondary targets could be placed for irradiation in an experiment.  

Preliminary 2-dimensional particle-in-cell simulations are consistent with these estimates of the reaction volume, its dimensions and confinement time.  The simulations were less successful in explaining the deuteron temperature and somewhat underestimated the fusion neutron yield.  These errors may arise from a combination of the 2-d simulation volume which does not resolve the 3-d dynamics of relativistic transparency and break-out after-burner \cite{yin20113dboasims} and the limited sampling of distribution functions inherent to the PIC algorithm.  The simualtions' agreement in reaction volume and confinement time supports the estimated neutron flux, which otherwise would be very sensitive to these parameters.

\section{Discussion}

Many groups have conducted experiments with various types of deuterated targets, ranging from deuterated plastic (CD$_2$) foil targets \cite{willingale2011comparison}, to deuteron cluster targets \cite{buersgens2006angular} and heavy water droplets \cite{karsch2003high}. The fusion neutron yield observed in this experiment using pure cryogenic deuterium on a petawatt class laser is two orders of magnitudes greater than in previous experiments. Several effects, including ion density, electron density and target thickness, contribute to this high neutron yield. 

The cryogenic deuterium target has advantages over other deuterated targets in its ion and electron densities.  The deuterons are uniformly distributed through target in contrast to deuterated clusters \cite{buersgens2006angular}, in which the large distances  between clusters of 1-10\si{\nano\meter} diameter limit reactions to within each cluster and reduce the overall number of deuterons likely to react. For CD$_2$ \cite{willingale2011comparison} or heavy water droplet \cite{karsch2003high} targets, the deuteron ion density is within a factor 2 of the solid-deuterium ion density, but the presence of other ions (C,O) reduces the fraction of laser energy transfered to the deuterons.  The electron density in the deuterium ice targets is five times lower. Lower electron densities reduces the gamma factor $\gamma\simeq \sqrt{1+a_0^2/2}$ ($a_0$ is the normalized vector potential) required to achieve relativistic transparency by five times and hence reduces the required laser intensity by 25 times.  Relativistic transparency facilitates volumetric heating, a larger reaction volume and less laser energy reflected \cite{hegelich2013laser}. Smaller gamma factor also implies lighter electrons; thus, they can be more easily expelled by the radiation pressure (ponderomotive force) \cite{macchi2013ion}. Without the screening of electrons, ions can directly extract energy from the laser field, which calculations suggest may account for ion kinetic energies up to 50 keV. As most of the laser energy is absorbed by the electrons in usual laser plasma interaction, a lower electron density also means less laser energy loss to electrons. 

For a given laser intensity and target density, a well-defined range of target thickness optimizes transfer of laser energy into the target plasma \cite{Luedtke:2020qgx}.  Targets that are too thin (hundreds of nanometers as in the case of plastic foils) expand after ionization, significiantly decreasing the ion density and have a limited reaction volume, leading to a reduced neutron yield. Hot electrons in thin targets can readily escape from the plasma and transform their kinetic energy to coulomb potential energy stored in the plasma, which does not help thermalizing deuterons and also leads to a faster expansion. The plasma confinement time also decreases as the thickness decreases. On the other hand, overly thick targets are not heated volumetrically.  When the laser does not penetrate the full width of the target, relativistically transparent plasma can become opaque again due to hole boring effect \cite{iwata2018plasma}. Cold electrons from unheated regions flow back into the interaction region where hot electrons are escaping. The cold electrons cool the plasma and screen the laser field, thus reducing the efficiency of energy transfer from laser to ions. All result in lower ion temperatures. Deuterium ice targets allow for thicker targets and hence larger reaction volume because their lower electron density decreases the threshold for relativistic transparency.

The most important reason is the fact that the fusion rate scales with $T^3$ and therefore is very sensitive to the plasma temperature in the tens of keV range. The plasma temperature inferred from the neutron spectrum was 200$\pm 50$ keV, which under thermal equilibrium conditions implies 3-4 orders of magnitude increase in the D-D fusion cross sections compared to ICF experiments (ion temperature $T_i\simeq 5$\si{\kilo\electronvolt} at NIF \cite{park2014high} and OMEGA \cite{hohenberger2012inertial}), cluster targets ($T_i\simeq 10$\si{\kilo\electronvolt} \cite{bang2013temperature}) or CD$_2$ targets.  Even lower estimates of the deuteron temperature achieved, $T_d\simeq 150$ keV, imply 2 orders of magnitude enhancement in reaction rate.  Although we do not expect the deuteron distribution to thermalize within the 10 ps confinement time, any approximately Maxwellian distribution or combination of Maxwellian distributions for subsets of the deuteron population will ensure the neutron spectrum is fit by a single (effective) temperature parameter describing the deuteron distribution. 

\section{Conclusion}
We have demonstrated laser-driven high temperature D-D nuclear fusion reactions using novel solid targets providing high neutron yield ($10^8$/J of laser energy) and high peak flux ($>10^{25}$\si{\per\square\cm\per\second}) on the plasma surface. This peak neutron flux translates to a flux $>10^{19}$\si{\per\square\cm\per\second} on a surface 1\si{\centi\meter} away and is at least one hundred times higher than the laser-ion driven method ($10^{17}$ \si{\per\square\cm\per\second}) \cite{guler2016neutron} as well as conventional neutron source like spallation sources ($10^{17}$\si{\per\square\cm\per\second}) and fission reactors ($10^{15}$\si{\per\square\cm\per\second}) \cite{taylor2007route}. The greater fluxes achieved in this experiment are due to the small source size ($\sim (10\si{\micro\meter})^3$) and short source duration ($\sim 10\si{\pico\second}$). If harnessed, such a high peak flux is a step toward experiments previously infeasible, for example, the study of r-process fusion\cite{bartlett2006two}, responsible for the creation of heavy elements. The key requirement for this research is an extremely high flux of neutrons to allow successive neutron absorption at a rate faster than the decay time. A future multi-beam facility could irradiate a target of interest using two or more laser-driven fusion neutron sources with ultrahigh neutron flux at precision intervals of picosecond-scale.  

We have also created the highest fusion temperature (200 keV) in the laboratory, one order of magnitude higher than other methods.  The inferred fusion temperature is consistent between the fit to the fusion neutron spectrum and the total neutron yield.  The GEANT4 simulation also confirmed our analysis of the deuteron break-up reaction on the beryllium converter or the RCF stack, except for the discrepancy on the exact spectrum profile shape.  Our analysis shows that, particlarly with additional instrumentation and more precise measurements, neutrons and their originating nuclear reactions can be probe local temperature and density conditions \cite{yuan2005shock} in high-intensity laser experiments, which are inaccessible to conventional optical probes due to the plasma density and timescale of the interaction. This method has potential to significantly increase the peak neutron flux with the next generation lasers such as 10PW Extreme Light Infrastructure \cite{gales2018extreme} and multi-PW Apollon laser \cite{papadopoulos2016apollon}. These lasers have the potential to create plasma with larger reaction volume and higher fusion temperature, producing an even brighter neutron source in the near future.  The higher plasma densities and neutron fluxes achieved here provide another step toward laboratory study of extreme astrophysical events \cite{thielemann2011astrophysical}.

\begin{acknowledgments}
Work performed under the auspices of the University of Texas at Austin. This work was supported by the Air Force Office of Scientific Research (FA9550-14-1-0045). High performance computing resources were provided by the Texas Advanced Computing Center. This work used the Extreme Science and Engineering Discovery Environment (XSEDE), which is supported by National Science Foundation grant number ACI-1548562.  We would like to thank the Texas Petawatt laser facility staff for their brilliant and unwavering support.
\end{acknowledgments}

\begin{appendix}
\section{Supplementary experimental information}
RCF stack formula (from front to back): 13\si{\micro\meter} Al + HDv2 + 8$\times$(100\si{\micro\meter} Al +HDv2) + 6$\times$(150\si{\micro\meter} Cu + MDv3) + 16$\times$(500\si{\micro\meter} Cu + EBT3) + 5$\times$(1\si{\milli\meter} Cu + EBT3). \\
HDv2: Mylar, thickness:~105\si{\micro\meter}\\
MDv3: Mylar, thickness:~260\si{\micro\meter}\\
EBT3: Mylar, thickness:~280\si{\micro\meter} \\

\begin{figure}[h]
\includegraphics[width=0.7\textwidth]{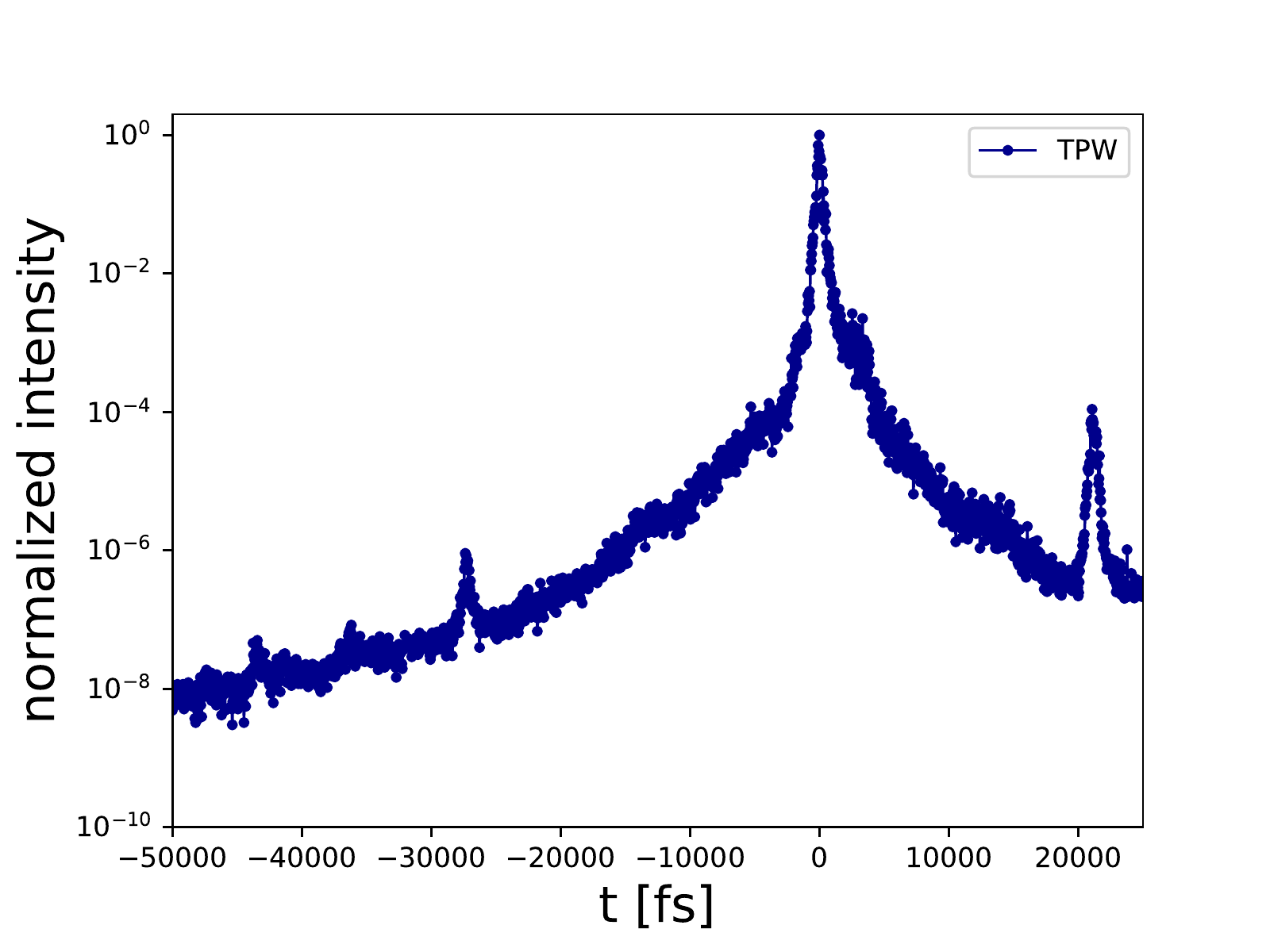}
\caption{Second-order autocorrelator measurement of the Texas Petawatt laser pulse.
\label{fig:TPWprofile}}
\end{figure}
\end{appendix}

\bibliographystyle{apsrev4-1}
\bibliography{fusion}

\end{document}